\documentclass[aps,pre,onecolumn,superscriptaddress,groupedaddress]{revtex4}

\usepackage{graphicx}
\usepackage{pgfplots}
\usepackage{color}
\usepackage{amsmath}
\usepackage{wasysym}
\usepackage{amssymb}
\usepackage{gensymb}
\usepackage{soul}
\usepackage{ulem}

\usepackage{algorithm}
\usepackage{mathtools}
\usepackage{verbatim}
\usepackage{subcaption}
\newcommand{\be}{\begin{equation}}
\newcommand{\ee}{\end{equation}}
\newcommand{\bea}{\begin{eqnarray}}
\newcommand{\eea}{\end{eqnarray}}
\newcommand{\bse}{\begin{subequations}}
\newcommand{\ese}{\end{subequations}}

\newtheorem{theorem}{Theorem}

\newtheorem{propn}[theorem]{Proposition}

\def\Pr{\mathbb{P}}
\def\bM{\mathbf M}
\def\bD{\mathbf D}
\def\bd{d_1^N}
\def\hm{\hat{m}}
\def\hM{\hat{\mathbf{M}}}

\makeatletter
\newcommand*{\rom}[1]{\expandafter\@slowromancap\romannumeral #1@}
\makeatother

\begin{document}
\title{High fidelity epigenetic inheritance: Information theoretic model predicts $k$-threshold filling of histone modifications post replication}




\author{Nithya Ramakrishnan}
\email{nithya.r@iitb.ac.in}
\affiliation{Department of Biosciences and Bioengineering, Indian Institute of Technology Bombay, Mumbai, India}
\author{Sibi Raj B Pillai} 
\email{bsraj@ee.iitb.ac.in}
\affiliation{Department of Electrical Engineering, Indian Institute of Technology Bombay, Mumbai, India}
\author{Ranjith Padinhateeri}
\email{ranjithp@iitb.ac.in}
\affiliation{Department of Biosciences and Bioengineering, Indian Institute of Technology Bombay, Mumbai, India}

%

%

\date{\today}
\begin{abstract}

Beyond the genetic code, there is another layer of information encoded as chemical modifications on histone proteins positioned along the DNA. Maintaining these modifications is crucial for survival and identity of cells. How the information encoded in the histone marks gets inherited, given that only half the parental nucleosomes are transferred to each daughter chromatin, is a puzzle. We address this problem using ideas from Information theory and understanding from recent biological experiments. Mapping the replication and reconstruction of modifications to equivalent problems in communication, we ask how well an enzyme-machinery can recover information, if they were ideal computing machines. Studying a parameter regime where realistic enzymes can function, our analysis predicts that, pragmatically, enzymes may implement a threshold$-k$ filling algorithm which derives from maximum \`a posteriori probability decoding. Simulations using our method produce modification patterns similar to what is observed in recent experiments.

\end{abstract}

\maketitle

\noindent
Why do our skin cells behave very differently from our brain cells even though they have  the same genetic code?
The reason is, beyond the genetic code, there are multiple layers of information encoded by wrapping and folding DNA into chromatin with the help of many proteins~\cite{goldberg2007epigenetics,allis2007epigenetics}.
Most of the DNA is wrapped around octamers of histone proteins making chromatin essentially like a string of beads made of nucleosomes (DNA+histones)~\cite{kornberg1974chromatin,kornberg1999twenty,van2012chromatin}. Each nucleosome carries chemical modifications, like acetylations and methylations, forming a one-dimensional pattern of histone marks along the chromatin polymer contour~\cite{hiscode_allis,allis2007epigenetics}  (see Fig.~\ref{models}(\subref{schematic1}) top panel). This pattern encodes a crucial layer of information 
as described below.
 
Each nucleosome is constituted by four kinds of histone proteins, namely H3, H4, H2A and H2B~\cite{kornberg1974chromatin,luger1997crystal,alberts2002molecular}. The histone modifications are indicated by names like H4K5ac (acetylation modification on the $5^{\rm th}$ amino acid lysine (K) of the H4 protein), H3K27me3 (tri-methylation  marks on the $27^{\rm th}$ amino acid  lysine (K) of the H3 protein) etc~\cite{richards2002epigenetic,alberts2002molecular}.
 Presence of each of these modifications encode some specific information relevant for gene regulation. Even though the entire histone modification code is not deciphered yet, we understand it in parts. For example, H3K27me3 represses reading of the local DNA region where the modification is present, H3K9ac encodes for local gene activation and so on~\cite{kouzarides2007chromatin,alberts2002molecular,taverna2007long}. The activation and repression of genes collectively decide the gene expression pattern and hence determines the function and fate of a cell~\cite{rando2007global,weiner2015high,zentner2013regulation,seligson2005global}. 

While preparing to divide, cells copy their genetic code via the DNA replication process. For DNA to be copied, the chromatin has to be unfolded and histone proteins need to be disassembled~\cite{margueron2010chromatin,madamba2017inheritance}. This would disrupt the pattern of histone modifications. Recent studies have shown that the ${\rm{(H3-H4)}_2}$ tetramer from the parent remains intact~\cite{xu2010partitioning} and randomly -- with equal probability -- gets deposited onto either of the newly synthesised  DNA strands~\cite{groth2007chromatin,petryk2018mcm2,yu2018mechanism}. That is, daughter strands will have only some ($\approx 50\%$) of its nucleosomes from the parent; the rest need to be assembled from the pool of new histone proteins made afresh~\cite{alabert2015two,bameta2018coupling}. Since the new nucleosomes will not carry the histone marks present on the parental chromatin, half the parental marks are missing in each daughter chromatin~\cite{probst2009epigenetic}. 
  
Since the histone modification patterns decide the state (repressed/active) of all genes and the identity of the cell itself, it is crucial that the newly made chromatin re-establishes the pattern immediately after replication~\cite{reinberg2018chromatin}. Recent experiments show that many of the histone modification patterns --- patterns in the repressed gene regions, in particular --- are ``inherited'' from the parental chromatin~\cite{reveron2018accurate,escobar2019active}. How the new chromatin recovers the missing information and re-establishes a mother-like pattern is a puzzle.

It is known that there are specific enzymes to read and write histone marks~\cite{wang2004beyond}. While molecular details of some of these enzymes are known~\cite{sabari2017metabolic}, precisely what strategies they use to re-establish the histone modification pattern after replication are not fully understood. Even though, existing models discuss the establishment of modifications and other possibilities~\cite{zhang2015interplay,zheng2012total,fischle2003binary,alabert2020domain}, accurate re-establishment of a specific pattern is little explored. 

The problem of loss of information in the modification pattern during replication 
 and its retrieval within the daughter chromatin are very similar to data loss and 
 error correction in telecommunication.  In such systems, a transmitted signal gets exposed to noise and is consequently error-prone at the receiving end. The decoder at the receiver detects and corrects these errors using techniques from information theory and coding theory~\cite{madhow2008fundamentals,cover2012elements}. This viewpoint immediately poses the following questions: Can we use known decoding algorithms from communication theory to analyse chromatin modification loss and retrieval? How well can the best known algorithms correct the missing modifications and re-establish the modification patterns? What is the best possible correction strategy enzymes could use if they were ideal computing machines?
Are the algorithms compatible with biological processes that realistic cellular enzymes can conceivably do? In this paper, we address these questions using ideas from Information theory. We consider one of the daughter chromatins to be a noise-corrupted signal created at the replication fork, while the enzymes and other molecular agents help to correct this error using mathematical techniques. In this model, the inheritance of the mother's pattern is approached using Bayesian decoding techniques. Predictions from our model are verified
using publicly available experimental data, indicating the relevance of our work in studying real biological datasets~\cite{reveron2018accurate}.

\section*{Model and Methods}

\begin{figure}[tbhp]
 \begin{subfigure}[b]{.5\linewidth}
 \centering
\includegraphics[width=3in]{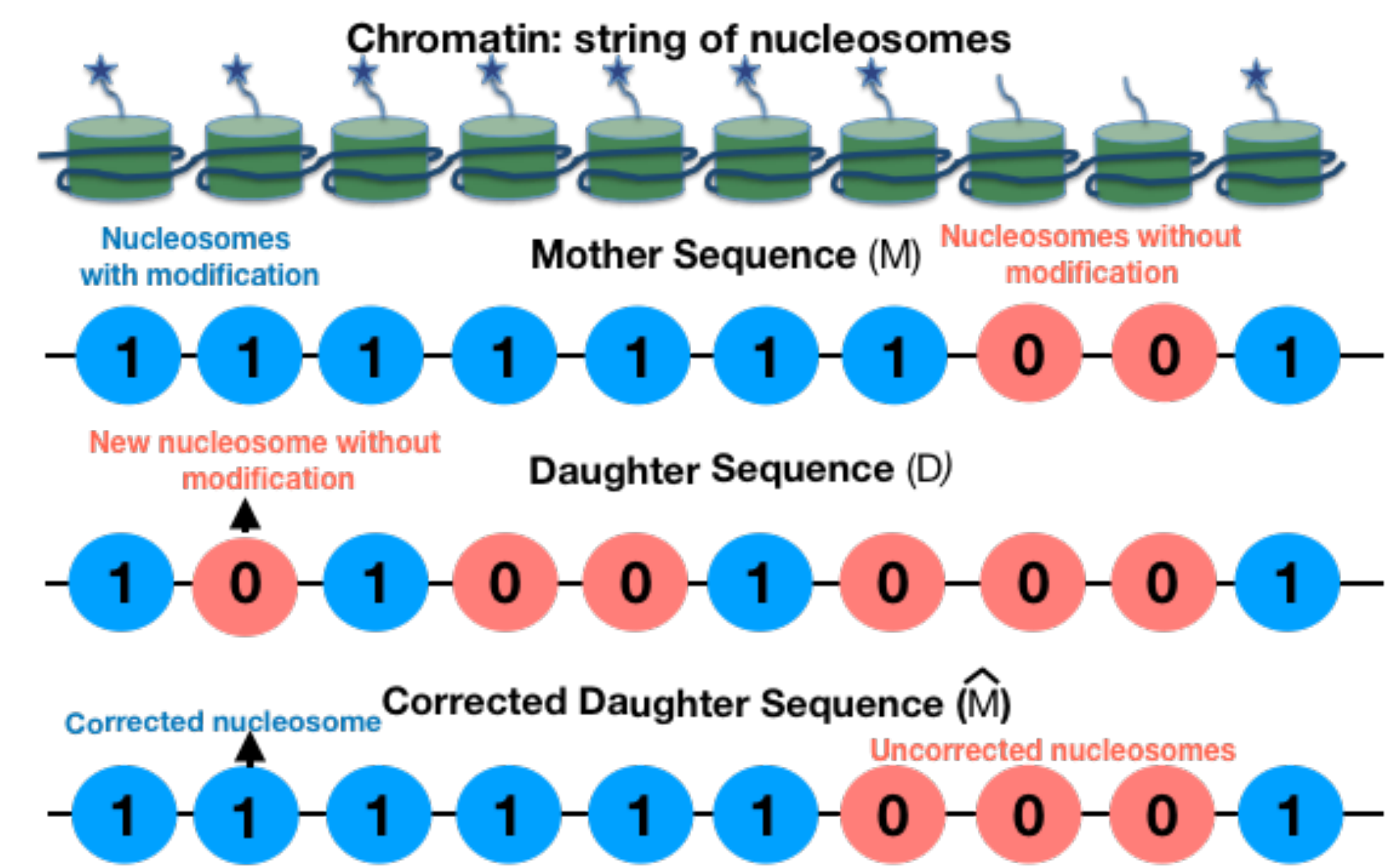}
\caption{} \label{schematic1}
  \end{subfigure}\vspace{0.3cm}

 \begin{subfigure}[b]{.5\linewidth}
 \centering
	\begin{tikzpicture}[line width=1.5pt, myline/.style={line width=1.0pt}]
\node (n0) at (0,0)[circle,draw,scale=1.5]{$0$};
\node (n1) at (4,0)[circle,draw,scale=1.5]{$1$};
\draw [->,myline] (n0) to[bend left=45] node[midway,below] {$1-\beta $} (n1);
\draw [->,myline] (n1) to[bend left=45] node[midway,above] {$1-\alpha $} (n0);
\path[->,myline] (n0) edge [loop/.style={in=150,out=210, looseness=8, min distance=10mm}, loop left] node[left]{$\beta$} (n0);
\path[->,myline] (n1) edge [loop/.style={in=-30,out=30, looseness=8, min distance=10mm}, loop left] node[right]{$\alpha$} (n1);
\end{tikzpicture}
\caption{} \label{markov_model_mother}
  \end{subfigure}\vspace{0.3cm}
 
 \begin{subfigure}[b]{.5\linewidth}
\centering
\begin{tikzpicture}[line width=1.5pt]
\node (m) at (0,0){$\bM$};
\node (z) at (2,-1.5){$ Z$};
\node (c) at (2,0)[circle,draw,scale=0.75]{AND};
\node (dec) at (4,0) [draw,thick, text width=1.25cm,minimum height=1.0cm, text centered] {Decoder};
\draw[->] (m) -- (c);
\draw[->] (z) -- (c);
	\draw[->] (c) -- node[above]{$\bD$}(dec);
	\draw[->] (dec) -- ++(1.5,0) node[right]{$\hat {\mathbf M}$};
\end{tikzpicture}
	\caption{} 
	\label{additive_noise}
  \end{subfigure}
  \caption{ Schematic description of the problem (a) Row 1 from top: chromatin as a string of nucleosomes with and without the histone modification(star) of our interest. This can be mapped to a string of binary numbers indicating the presence ($1$) or absence ($0$) of the modification (row 2), giving us $\bM$. Row 3: one typical realization of a daughter chromatin ($\bD$) produced from  $\bM$ above, via a process mimicking DNA replication, where only a fraction of the modifications ($1$s) will end up in the daughter chromatin, stochastically; the rest do not have the modification of interest ($0$).  Soon after replication, certain enzymes will insert modifications correcting $\bD$ to a mother-like sequence $\hat{\b{M}}$ (row 4). Since these are stochastic processes, we expect some errors. 
(b) The mother sequence ($\bM$) is modelled as a first order Markov chain having sequence of $0$s and $1$s. $\alpha$ and $\beta$ are probabilities of finding a $1$ followed by a $1$, and a $0$ followed by a $0$, respectively.  $1-\alpha$ and $1-\beta$ are probabilities of finding a $1$ followed by a $0$ (note arrowheads), and a $0$ followed by a $1$, respectively. 
(c) From an information theory perspective, the daughter sequence ($\bD$ ) is obtained by a mother sequence $\bM$ getting logically ANDed with an independent and identically distributed (IID) binary sequence $\mathbf{Z}$ (noise). A mother-like sequence $\hat{\mathbf M}$ is reconstructed by passing $\bD$ through a decoder. What are the plausible ways enzymes could act as decoders is the subject of this study.
} \label{models}
	
\end{figure}
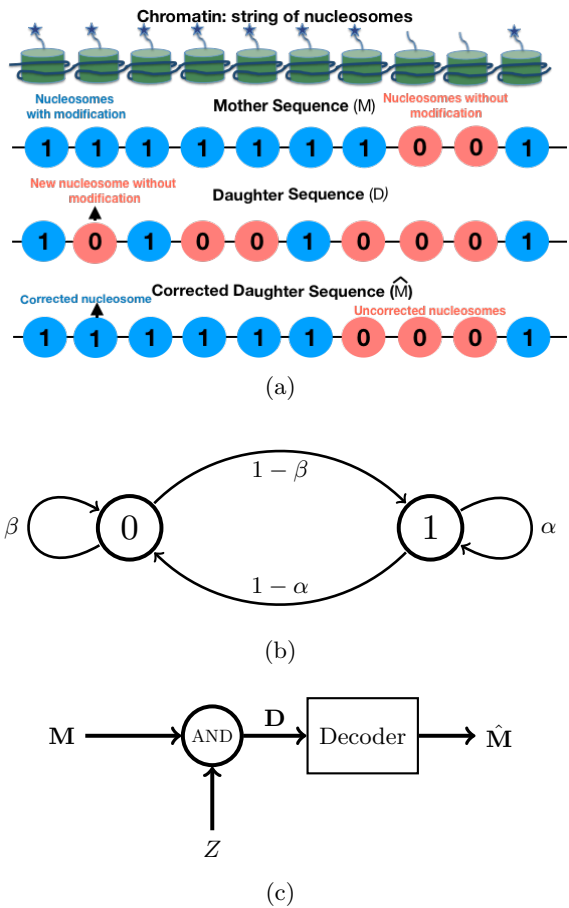

Consider a region on a mother chromatin having $N$ nucleosomes. We are interested in studying the inheritance of one histone modification at a time. Since many of the repressive marks are known to be inherited immediately after replication, we will consider one such repressive mark (e.g., H3K27me3) and its pattern along a chromatin. This pattern can be represented by a vector $\bM =\{ m_1, m_2, \cdots , m_N\}$, where $m_i$ can have values $1$ or $0$ indicating the presence or absence of the modification on the $i^{th}$ nucleosome (see Fig.~\ref{models}(\subref{schematic1})).
We also need the following notations:
\begin{itemize}
\item $m_i^j$ is a vector $(m_i, m_{i+1},\cdots, m_j)$ with $j> i$; the subscript and superscript respectively indicate the first and last indices. Thus,  $\bM =  m_1^N$ is a realisation of the entire mother chromatin modification sequence.

\item a row vector of $k$ consecutive ones (zeroes) will be denoted as $1_{k}$ ($0_k$). Extending this, 
the vector $(1,0,\cdots,0,1)$ representing an island of $k$ consecutive zeroes between two 
ones will be denoted as $(1,0_k,1)$.
\end{itemize}
Since modification on a nucleosome is very likely related to its immediate neighbours, we model the pattern $\bM$ along the mother chromatin as a binary valued
random walk, having neighbourhood interactions corresponding to a first order homogenous Markov chain. More specifically, given the modifications $m_{i-1}$ and $m_{i+1}$, the modification $m_i$ is assumed to be
independent of all other modification values.
Equivalently, the conditional probability law is
\begin{align} \label{eq:markov:chain}
\Pr(m_i|m_1, \cdots, m_{i-1}) = \Pr(m_i|m_{i-1})\,,\,\, i \geq 2,
\end{align}
where $m_1$ is the modification on the first nucleosome of the region of our interest. The  state-space evolution of the Markov chain $\bM$ is as follows:
given $m_i=1$, let $\alpha$ and $1-\alpha$ be the probabilities for obtaining $m_{i+1}=1$ and $m_{i+1}=0$ respectively. Similarly, if $m_i=0$, let $\beta$ and $1-\beta$ be the probabilities for having $m_{i+1}=0$ and $m_{i+1}=1$ respectively. 
The sequence $\bM$ can be seen as a random walk on the state space shown in Fig.~\ref{models}(\subref{markov_model_mother}). For example, 
when $\alpha$ and $\beta$ are close to $1$, the pattern would often contain long runs of either $1$s (presence of modification) or $0$s (absence of modification).


From the mother chromatin $\bM$, the generation of a daughter chromatin having 
histone  modification sequence $\bD = \bd$  is modelled as follows. 
During replication, with probability $\frac{1}{2}$, each nucleosome on a daughter chromatin is either directly inherited from its parental counterpart (i.e. $d_i=m_i$) or newly deposited (i.e. $d_i=0$) from a pool of fresh histones assembled de novo~\cite{probst2009epigenetic}. 
This process is equivalent to doing a logical AND operation of the mother sequence \textbf{M}  with an independent and identically distributed (IID)  binary sequence $\mathbf{Z}$ (noise), i.e. $\bD = \bM.\mathbf{Z}$ (see Fig.~\ref{models}(\subref{additive_noise})). This  biological process leads to a memoryless model with conditional probability
\begin{align} \label{eq:dmc}
	\Pr(d_1^N|m_1^N) = \prod_{i=1}^N \Pr(d_i|m_i).
\end{align}
In biology, the question is, given a daughter sequence $\bd$ soon after replication, how could a cell reconstruct  a mother-like sequence  $\hat{{\bf M}}=\hat{m}_1^N$? In other words, the question is how to find a decoder that would reconstruct a $\hat{{\bf M}}$  from ${\bf D}$ (see Fig.~\ref{models}(\subref{additive_noise})). Ideally, a cell would want to choose a binary sequence $\hM$ having the minimum deviation from $\bM$. The fraction of errors in the reconstructed  sequence is a highly desired deviation metric, given by
\begin{align} 
\label{eq:object}
	\Delta(\bM,\hM) = \frac{1}{N} \sum_{i=1}^N (m_i - \hat{m}_i)^2.
\end{align}
This deviation metric is effectively the bit error rate (BER) when $N$ 
becomes large~\cite{LinCostello}.
%
%
Thus the chosen $\hM$ should
minimize the BER with respect to the actual sequence $\bM$, while obeying the transition
law in Eq.~\eqref{eq:markov:chain}.  This is similar to data communication through an erroneous channel.
It is well known that  Bayesian estimation schemes minimize the average detection error probability at the receiver. In particular, a decoder choosing the input sequence having the Maximum \`Aposteriori Probability (MAP) is optimal in minimising the communication error~\cite{LinCostello,cover2012elements}. We call this the Sequence MAP (SMAP) decoder, which 
 identifies the most probable sequence $\hM=(\hm_1, \cdots, \hm_N)$ based on
the observations $d_1^N$ as
\begin{align} \label{eq:seq:map:0}
	(\hm_1, \cdots, \hm_N) = 
	\underset{m_1,\cdots,m_N}{\text{argmax}}
	\Pr (m_1^N|\bd).
\end{align}
SMAP decoding is known to have near optimal BER performance, and good analytical tractability in many 
contexts~\cite{LinCostello}. While the optimal BER performance can be achieved by
Bitwise MAP (BMAP) decoding for each modification value separately,
the latter scheme is not only computationally
more demanding, but also analytically less tractable. It appears that
SMAP decoding is a potential candidate for biological cells to reconstruct 
the epigenetic modifications from the partial data. 
Notice that SMAP decoding depends on the parameters $\alpha$ and $\beta$ of the Markov Chain. Clearly, this algorithm  is only targeting a primary reconstruction immediately following DNA replication; secondary mechanisms may further alter the patterns in the long run~\cite{groth2007chromatin,bonasio2010molecular,allshire2018ten}.


\section*{Results}
In this section, we will be using the ideas developed in the Model and Methods to answer how one can reconstruct a mother-like modification sequence ($\hM$), given the daughter chromatin sequence $\bD$. We will discuss how well algorithms like the Sequence MAP (SMAP) decoding will compute $\hM$, and whether realistic enzymes can implement this in practice.


\subsection*{Ideal Enzymes Implementing SMAP Decoding}
While we do not know exactly how biological enzymes work to retrieve histone modification patterns soon after replication, how well can the communication theory-inspired algorithms reconstruct a mother-like sequence? To test this, one can imagine some \emph{ideal enzymes}---computing machines---constructed to implement the SMAP algorithm in Eq.~\eqref{eq:seq:map:0}
That is, these enzymes will maximize the conditional probability $\Pr(\hM|\bD)$ based on the given daughter sequence observations $\bD=d_1^n$. 
Applying Bayes' rule ~\cite{stewart2009probability}, along with Eq.~\eqref{eq:markov:chain} and Eq.~\eqref{eq:dmc}, one gets
\begin{align} \label{eq:seq:map:1}
	\Pr(\hM|\bD) = \frac 1{\Pr(\bD)}\prod_{i=1}^N \Pr(\hm_i|\hm_{i-1}) \Pr(d_i|\hm_i).
\end{align}
Since $\Pr(\bD)$ does not play a role in the maximization over $\hM$, it can be ignored for our purposes.  
Ideal computing machines can now implement the SMAP algorithm using the idea of trellis decoding~\cite{sklar2001digital}, which is closely 
related to the well known Viterbi Algorithm in coding theory~\cite{LinCostello}.
While the memory and computational power requirement for trellis decoding is high in general, we find that decoding procedure for our model can be broken down to 
smaller sub-sequences, each corresponding to a different run of zeroes in $\bD$. 
In particular, trellis decoding has to be applied only on sub-sequences of the form
$(1, 0_k, 1)$, which has $k$ consecutive zeros in between two ones. The following proposition verifies
this claim.
%
\begin{propn}
	Let $i,j$ be two positions where the daughter sequence has ones, with $j>i$. 
	Then
	$$
	\Pr (m_i,\cdots, m_j|\bd, m_1^{i-1}, m_{j+1}^N) = \Pr (m_i,\cdots, m_j|d_i^j).
	$$
\end{propn}

To prove this, the following chain of equalities are valid under
	the assumption that $(m_i,m_j) = (d_i,d_j) = (1,1)$.
\begin{align*}
	\Pr (m_{i+1},\cdots, m_{j-1}|d_i^j) &= \Pr (m_{i+1},\cdots, m_{j-1}|d_i^j, m_i,m_j) \\
	&= \Pr(m_i^j|d_i^j, m_i,m_j, m_1^{i-1},m_{j+1}^N) \\
	&= \Pr(m_i^j|d_i^j, m_1^{i-1},m_{j+1}^N, d_1^{i-1},d_{j+1}^N) \\
	&= \Pr(m_i^j|\bd, m_1^{i-1}, m_{j+1}^N). 
\end{align*}
	Notice that the second step used the Markov assumption 
	that given the current state, the future and past states are 
	conditionally independent. The third step used the facts that $(m_i,m_j) = (d_i,d_j)$,
	and $(d_1^{i-1},d_{j+1}^N)$ is generated from 
	$(m_1^{i-1},m_{j+1}^N)$ using random variables independent of $(m_i^j, d_i^j)$.


Thus, while performing SMAP decoding by Eq.~\eqref{eq:seq:map:1}, the decisions on $\hm_i, \cdots, \hm_j$ 
are independent of the values $(d_1^{i-1}, d_{j+1}^n)$, once $d_i=d_j=1$.  In other words,
to decide on a bit  at
position $l$ where the daughter has inherited a zero,
we need to only consider the smallest daughter segment containing the position~$l$,
and flanked by ones at both ends.
Only daughter segments with at least one intermediate zero are to be considered;
otherwise there is nothing to decode. Without loss of generality, we will
take $\bD=(1,0_{k}, 1)$ for the rest of the exposition, corresponding
to a run of $k$ zeros, and perform trellis decoding on this sequence.

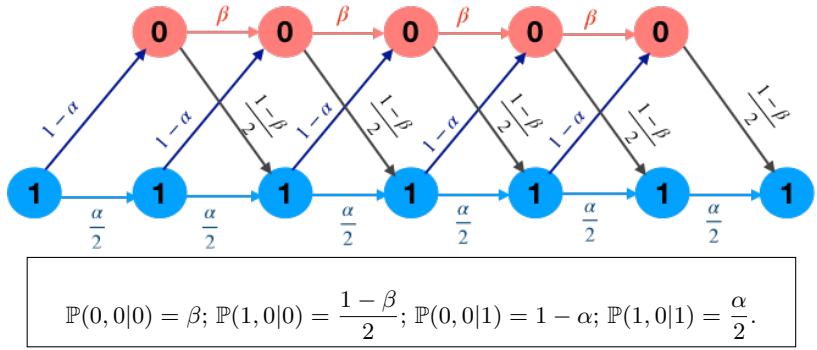
\begin{figure}[tbhp]
\centering
\includegraphics[width=0.6\linewidth]{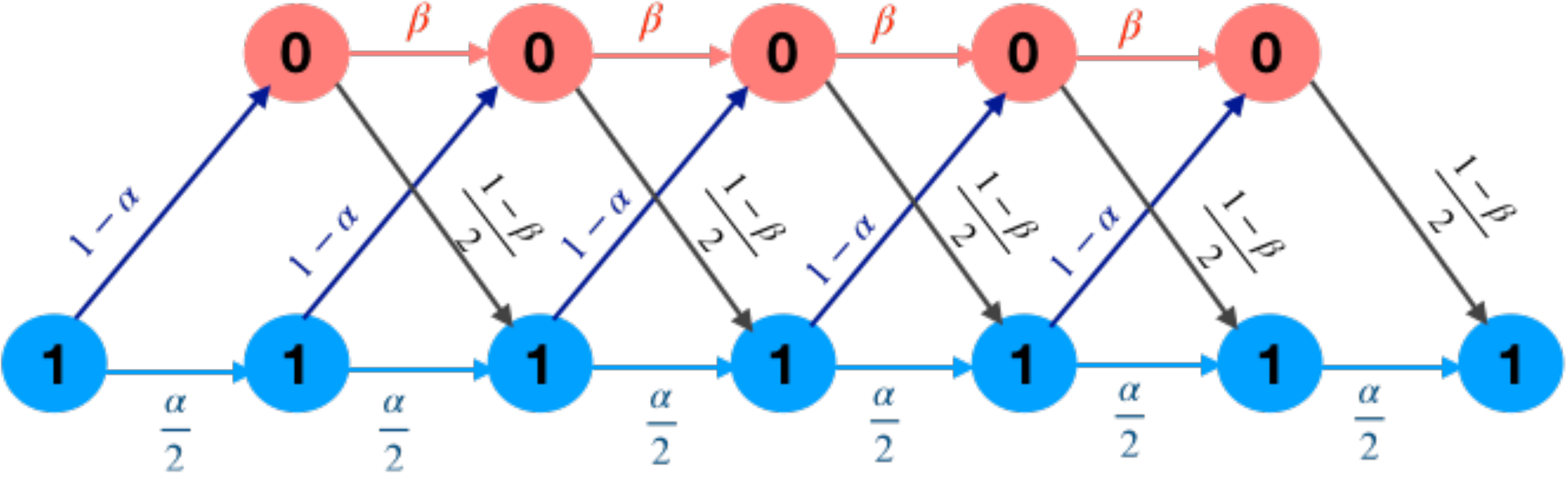}
\begin{tikzpicture}

	\node at (-0.1,-2.1)[right, text width=10cm, rectangle, draw,thin]{\small
		\begin{align*}
			\Pr(0,0|0) = \beta; \hspace{0.5mm}
			\Pr(1,0|0) = \frac{1-\beta}2 ;\hspace{0.5mm}
			\Pr(0,0|1) =  1-\alpha ;\hspace{0.5mm}
			\Pr(1,0|1) = \frac {\alpha}2.
		\end{align*}
		};

\end{tikzpicture}
\caption{The trellis diagram used for illustrating how the MAP algorithm chooses the path of maximum a posteriori probability given a daughter sequence $(1,0_5,1)$. From each Markov state ($0$ or $1$), there are two possible arrows to transition to the next state. Given that the daughter state $d_i=0$, all possible values $\Pr(m_i,d_i|m_{i-1})$ are given in the box, and represent probabilities associated with each arrow.  Starting from state $1$ at the left bottom, one can choose all possible paths moving along the arrows, and compute the path probabilities (path metrics) using Eq.~\eqref{eq:seq:map:1}. We choose the path with the largest path metric.}
\label{fig:viterbi}
\end{figure}

{\bf Trellis Decoding: } In a trellis, the idea is to find probabilities of all possible ``paths'' ---sequence of states of the Markov chain. Given that our interest is to examine daughter sequence regions like $(1,0_k,1)$, Fig.~\ref{fig:viterbi} demonstrates the trellis diagram for a run of $5$ zeroes ($k=5$). Starting from the initial state $1$ (left bottom in Fig.~\ref{fig:viterbi}), the trellis diagram assigns a conditional probability (branch metric) to each subsequent 
transition (arrow), based on the  transition probability law and observed daughter state. 
For transitions from state at $i-1$ to $i$, 
we mark the branch metrics as $\Pr(m_i|m_{i-1})\Pr(d_i|m_i)$, where $d_i=0$ for $2 \leq i \leq N-1$. While the sequence $\bD$ is easily seen to be generated by a hidden Markov Model (HMM),  the branch probability metrics can be understood from the box shown in Fig.~\ref{fig:viterbi}. Notice that we have to find $\Pr(m_i,d_i=0|m_{i-1})$ for $(m_{i-1},m_i) \in \{(0,0),(0,1),(1,0),(1,1)\}$ (see Fig.~\ref{fig:viterbi}).

Notice that each possible mother sequence can be identified as a path in the trellis, 
with the labels identifying the branch metrics.  Using Eq.~\eqref{eq:seq:map:1}, 
the product of  corresponding branch metrics will yield the path metric  of each possible sequence, and then  the path maximizing the SMAP metric can be chosen. 
Since $\bD$ has the form $(1,0,\cdots, 0,1)$,  the start and end states of the trellis are ones (left and right bottom in Fig.~\ref{fig:viterbi}).

\begin{figure}[tbhp]
\centering
\includegraphics[width=.6\linewidth]{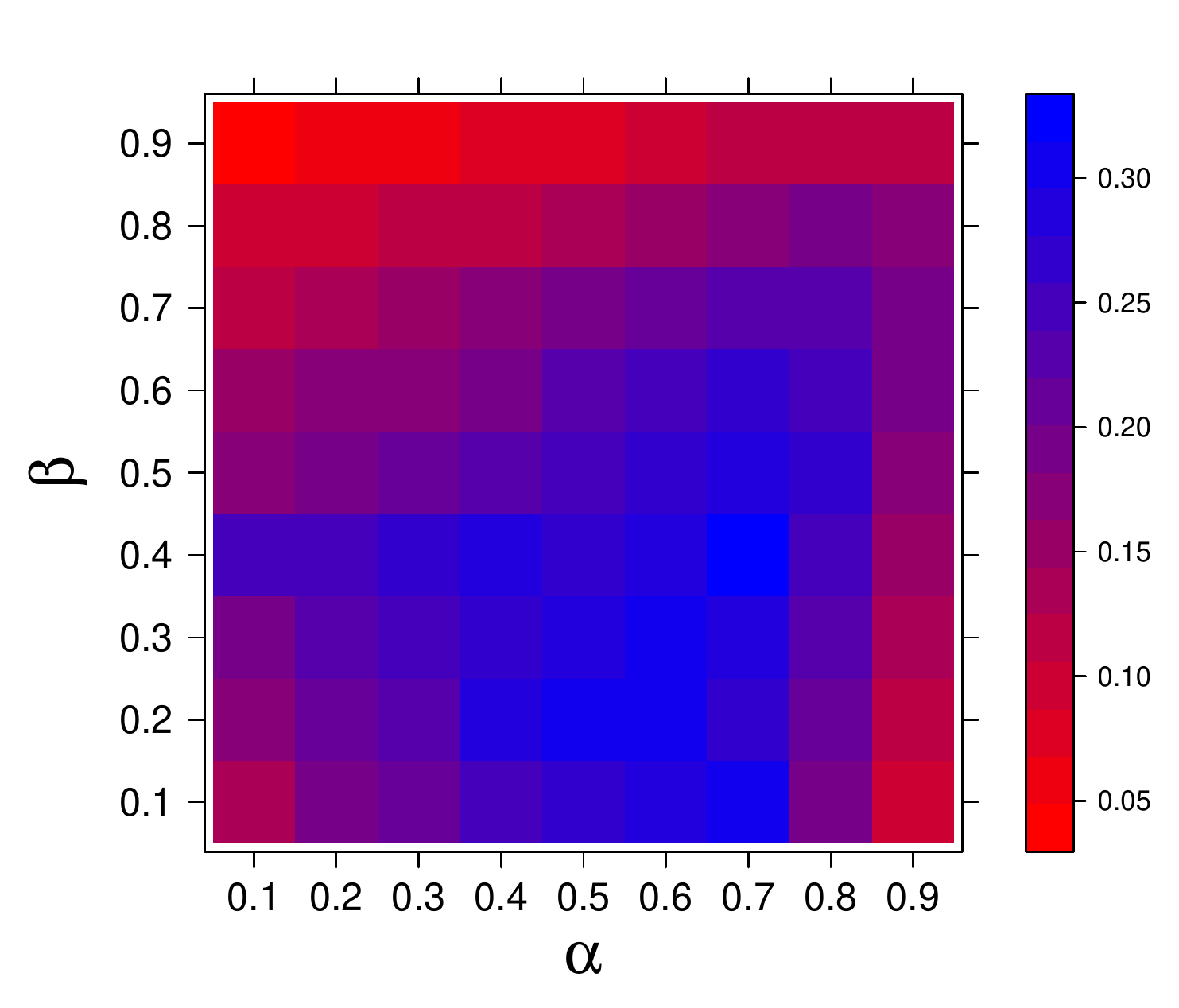}
\caption{The average deviation between the original mother and the mother-like corrected daughter sequences ($\bar{\Delta}=$ ensemble averaged $\Delta (M,M^{\prime})$) is plotted for different $\alpha$ and $\beta$ values as a heatmap (see color bar on the side).  The error is averaged over the error 300 mother sequences and 200 daughter sequences corresponding to each mother sequence.
\label{fig:Heatmap_ab}
}
\end{figure}

On a computer, we generated several mother sequences for different values of $\alpha$ and $\beta$ parameters in the Markov model. We then generated several daughter sequences, for each of the mother sequences, by flipping $1$s to $0$s randomly with probability $0.5$; that is, ${\bf D}=\bf{M} \cdot \bf{Z}$, with $\bf{Z}$ an IID binary sequence generated by an unbiased coin. Each of the daughter sequences were corrected with the trellis-based SMAP decoding to generate the corresponding estimate $\hat{\bf M}$; the error 
$\Delta(\bM, \hM)$ was computed (Eq.~\eqref{eq:object}). The average of $\Delta(\bM,\hM)=\bar{\Delta}$ (averaged over many realizations) for fixed pairs of $\alpha$ and $\beta$, is presented as  a heatmap in  Fig.~\ref{fig:Heatmap_ab}.  The mean deviation($\bar{\Delta}$) between mother and corrected mother-like daughter is low for  very high values of $\alpha$ and $\beta$ --- a region dominated by long islands of ones (modified nucleosomes) separated by long islands of zeros (unmodified nucleosomes). There are other regions too where $\bar{\Delta}$ is relatively small. Overall, this result shows how well an 
ideal computing machine that employs state of the art information theory algorithms can recover the original mother sequence. The remaining question is, can
a real enzyme do as good as this computing algorithm?

\subsection*{Threshold-$k$ model: enzymes filling unmodified islands of size at most $k$ maintain chromatin fidelity}
Whether biological enzymes are equipped to do complex SMAP computations like trellis decoding by themselves is debatable. Nevertheless, we argue that in certain biologically relevant parameter regimes, the decoding rule can be simple enough for enzymes to potentially execute. 
Among the known histone modification patterns, it is common to have regions densely filled by a certain modification (e.g., H3K27me3), and other regions where the modification is totally absent. This corresponds to higher values
of $\alpha$ and $\beta$ in our Markov model. Below we show that in this regime, the SMAP algorithm simplifies to tasks that the enzymes may easily carry out.

 \begin{figure}
 \begin{subfigure}[b]{0.3\textwidth}
\includegraphics[width=\textwidth,height=0.2\textheight]{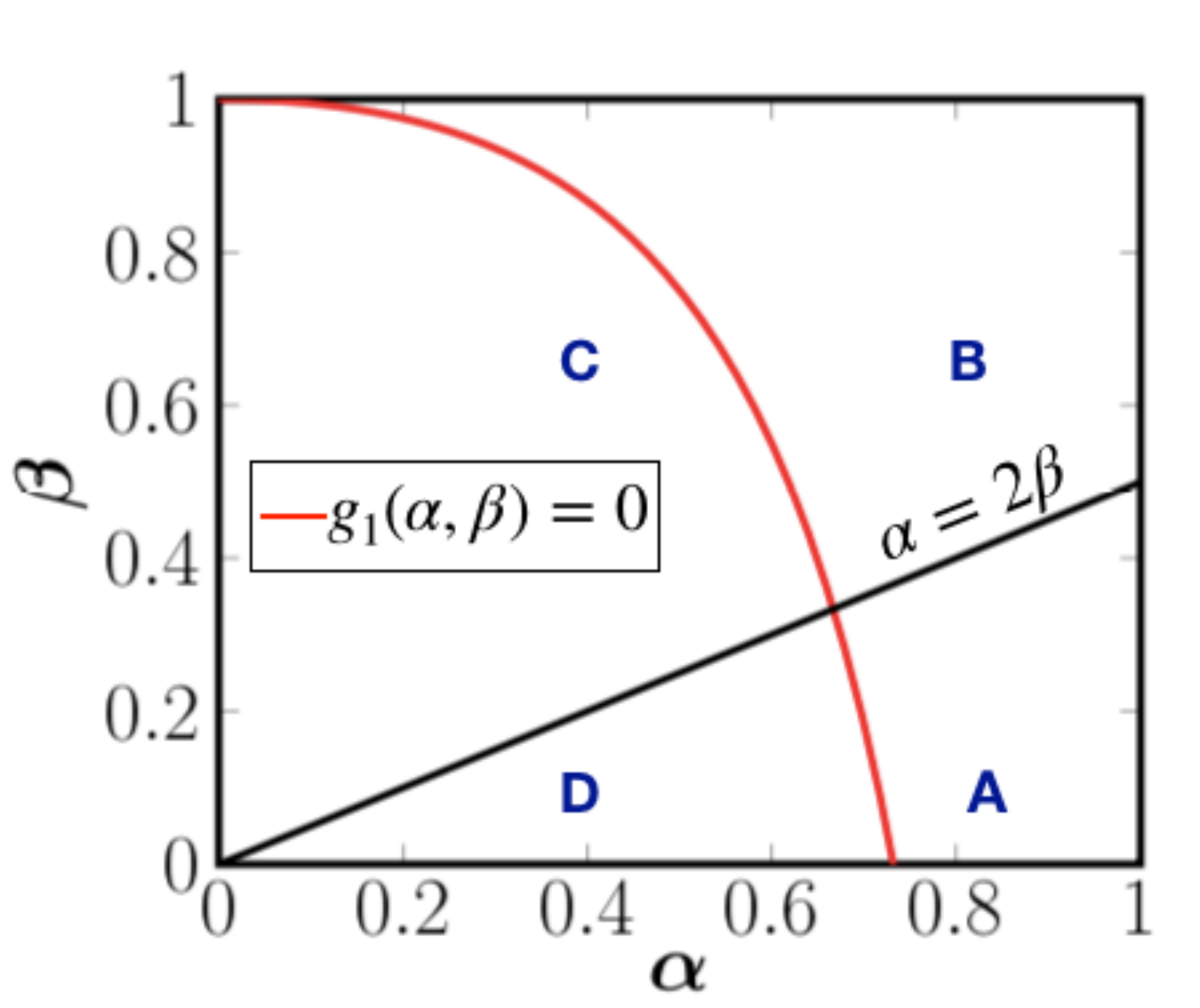}
\caption{} \label{smap_plot_region}
  \end{subfigure}  

  \begin{subfigure}[b]{0.3\textwidth}
\includegraphics[width=\textwidth]{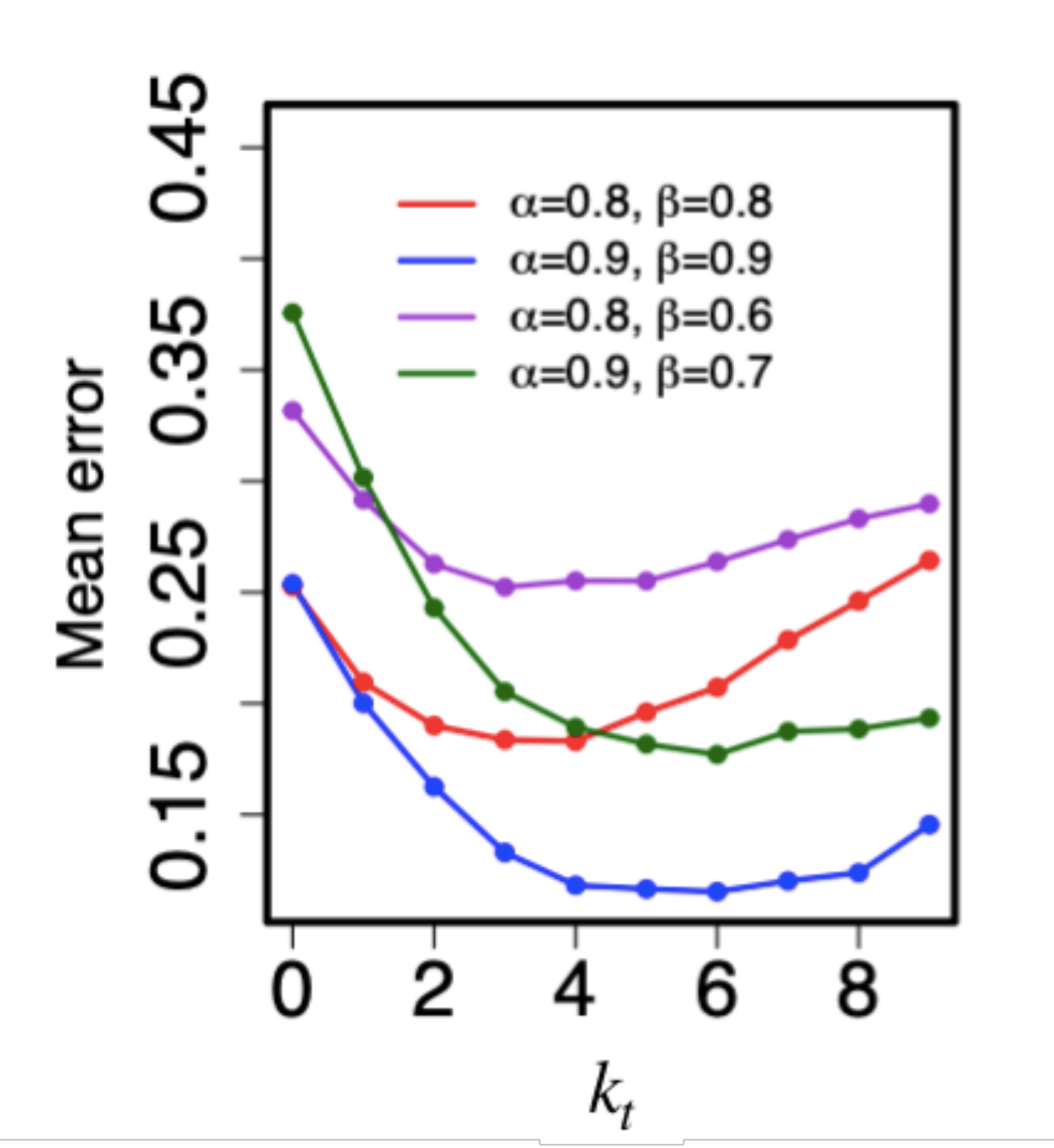}
\caption{} \label{opt_k_lineplot_b}
  \end{subfigure}  
     \begin{subfigure}[b]{0.3\textwidth}
\includegraphics[width=\textwidth]{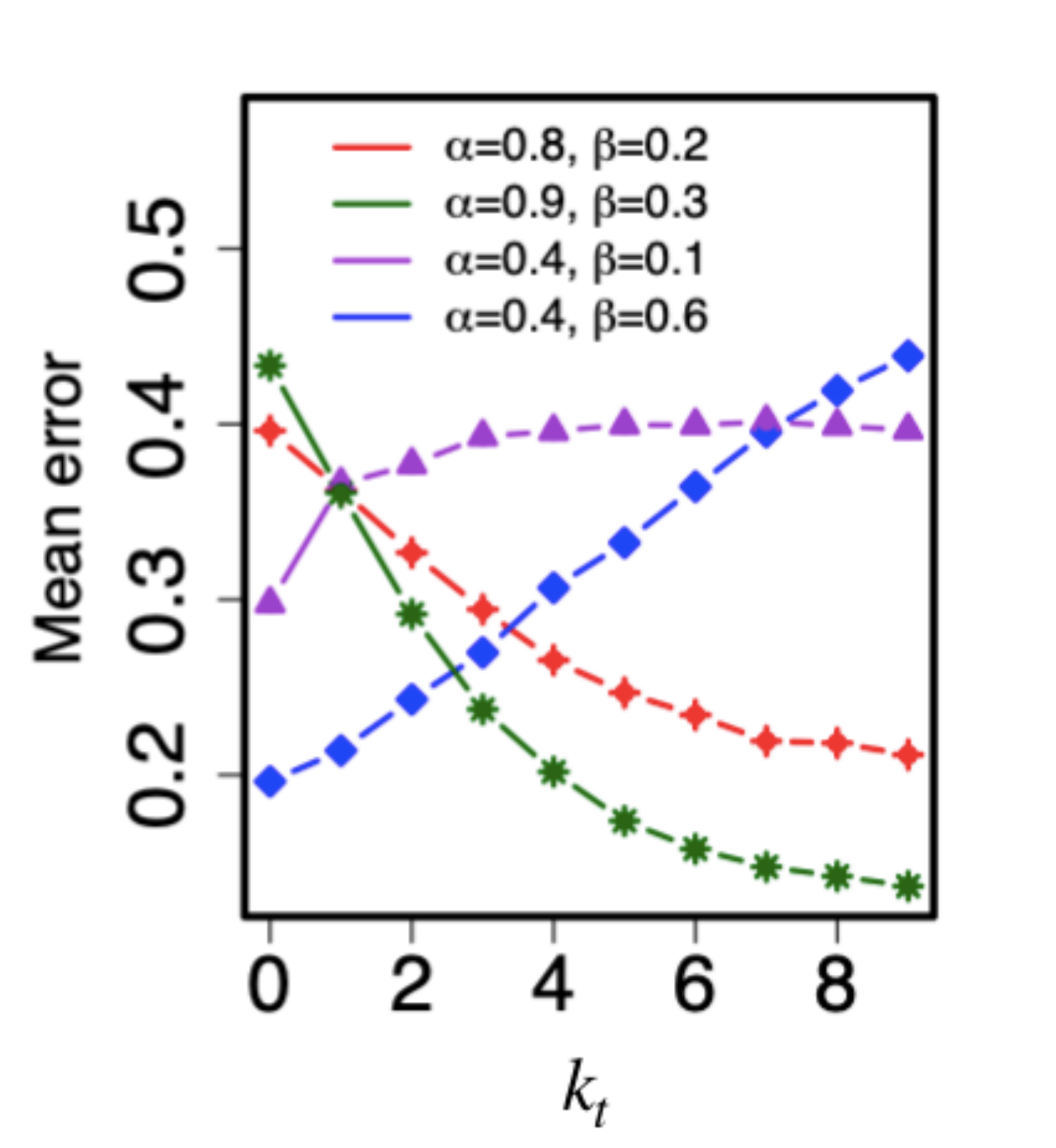}
\caption{} \label{opt_k_lineplot_c}
  \end{subfigure}

	\caption{ (a) Four (A, B, C, D) regions in $(\alpha, \beta)$ parameter space. The curves shown are
	$g_1(\alpha,\beta)=0$, and $\alpha-2\beta=0$.
	In region $A$ ($g_1>0, \alpha > 2\beta$), the  SMAP will replace every $0$ with $1$. In the region B ($g_1>0, \alpha < 2\beta$), enzymes can implement threshold-$k$ filling (see text). This parameter regime is realistic, biologically. \label{fig:filling_condition}
	 (b) The mean error ($\bar{\Delta}$) when we fill all islands of $0$s having size at most $k_t$. All these curves have $(\alpha,\beta)$ values in region B, and $\bar{\Delta}$ is non-monotonic having a finite optimum $k_t=k^*$. 
	 (c) $\bar{\Delta}$ for parameter values in region $A$ (red and green curves) are monotonically decreasing suggesting that the optimal $k_t$ is unbounded; hence the least error would be when all $0$s are replaced with $1$s. In regimes $C,D$ (blue and violet curves) the mean error is minimal when nothing is filled suggesting that threshold-$k$ filling is not suitable here.  Curves are obtained by averaging over $300$ mother, and 200 daughter sequences corresponding to each mother sequence. The standard error is smaller than the size of the points. }
\label{smap_optk}	
\end{figure}
%

Consider an island of $k$ unmodified nucleosomes in the
daughter chromatin, giving the pattern $(1,0_{k},1)$. From the trellis diagram (Fig.~\ref{fig:viterbi}), it can be seen that, if $\left(\frac\alpha2\right)^{k+1}>\frac 12 (1-\alpha) \beta^{k-1} (1-\beta)$, the probability of having the all-one path $(1,1_{k},1)$ at the mother is greater than that of ($1, 0_k, 1$). In other words, when the function $g_k(\alpha,\beta)
=\left(\frac\alpha2\right)^{k+1} - \frac 12 (1-\alpha) \beta^{k-1} (1-\beta)$ is positive,  an all-ones path is preferred over a run of $k$ zeros by SMAP decoding. 
However, the question is, for what values of $\alpha$ and $\beta$, does the above condition hold true. The expression
$g_k(\alpha,\beta)=0$ is easy to solve if we take $k$ to be a real value, this yields the root $k^*$ as:
\begin{equation}
k^*= \frac{\log \left( \frac{(1-\alpha) (1-\beta)}{(\alpha^2/2) }\right)}{\log (\alpha/2 \beta) }+1.
\label{eq:kstar}
\end{equation}
Notice that the solution for $k^*$ is unique when $0< \alpha < 1$ and $0<\beta<1$. 
Since $g_1(\alpha,\beta)=\frac 14 \alpha^2 - \frac 12 (1-\alpha)(1-\beta)$, $g_1(\alpha,\beta)>0$ implies that the sequence $(1,1,1)$ is preferred over $(1,0,1)$.
In addition, this condition will also imply that the numerator of Eq.~\eqref{eq:kstar} is positive.
The uniqueness of $k^*$ will now have the following implications:

On a unit square region of parameters $(\alpha,\beta) \in (0,1)\times (0,1)$, 
\begin{itemize}
\item When $\alpha < 2\beta$  and  $g_1(\alpha,\beta)>0$, one gets $k^*\geq 1$; we find that $g_k(\alpha,\beta)>0$ for all positive integers $k\leq k^*$ in Eq.~\eqref{eq:kstar}.
			This suggests that the SMAP algorithm will replace $(1,0_{k},1)$ with $(1,1_{k},1)$, if and only if $k \leq k^*$.
		
	\item When $\alpha > 2\beta$ and  $g_1(\alpha,\beta)>0$, we find that $g_k(\alpha,\beta)>0$ for any positive integer $k$; hence the SMAP algorithm will replace $(1,0_{k},1)$ with $(1,1_{k},1)$, for any value of $k$. 
%
%
\end{itemize}
Notice that when every path of at most $k^*$ zeros between two ones has less path metric than the corresponding all ones path, 
clearly any possible path other than all ones cannot have the maximum SMAP metric, while decoding sequences of length less than $k^*$.

The above analysis based on trellis decoding suggests two simple ways for enzymes to work. Enzymes of Type A would simply modify all unmodified nucleosomes ($0$s) between two modified nucleosomes ($1$s). Such enzymes may be preferred when the modification pattern can be modelled by parameters $\alpha$ and $\beta$ that corresponds to region $A$ in Fig.~\ref{smap_optk}(\subref{smap_plot_region}); notice that this has large $\alpha$ and small $\beta$. 
An enzyme of Type B would fill all unmodified nucleosomes ($0$s), if and only if the size of the unmodified region is $\leq k^*$. That is, replace $(1, 0_k,1)$ by $(1, 1_k,1)$, if the island size is $k\leq k^*$. Thus long islands of $0s$ are left unfilled. We call this a {\bf threshold-$k$ filling model}, which becomes active in region B of Fig. \ref{smap_optk}(\subref{smap_plot_region}).
Notice that when both $\alpha$ and $\beta$ are close to $1$, the modification is expected to have long domains (islands) with its presence, followed by islands with no modification. Biologically, this is a realistic regime for many modifications where enzymes can do threshold-$k$ filling.

We tested the threshold-$k$ filling model on a computer by generating several mother and daughter sequences, for various values of $\alpha$ and $\beta$; each daughter sequence was corrected using the threshold-$k$ filling algorithm --- that is, we filled all islands of $0$s, having size $k \leq k_t$, by 1s; here $k_t$ is taken as a variable. Corresponding mean error ($\bar{\Delta}$), averaged over many realisations, was computed for a given ($\alpha$, $\beta$). In Fig.~\ref{smap_optk}(\subref{opt_k_lineplot_b}), all the curves correspond to ($\alpha$, $\beta$) in regime B of Fig.~\ref{smap_optk}(\subref{smap_plot_region}) ($g_1>0$ and $\alpha < 2\beta$). In this interesting regime, the mean error has a non-monotonic behaviour with a minimum at a particular value of $k_t=k^*$, where $k^*$ is predicted by Eq.~\eqref{eq:kstar}. Note that $k^*$ values are around $3$ to $6$: enzyme of Type-B can fill small unmodified islands having  $3$ to $6$ zeroes, and leave much longer unmodified islands unfilled.
 
In Fig.~\ref{smap_optk}(\subref{opt_k_lineplot_c}), the two monotonically decreasing curves belong to the parameter regime $A$ ($g_1>0$, $\alpha > 2 \beta$). The mean error is decreasing as we increase $k_t$, suggesting that there is no finite $k^*$. If the modification patterns were in this regime, the corresponding enzymes should attempt to fill every unmodified region, however small or big that may be.  The other two curves in Fig.~\ref{smap_optk}(\subref{opt_k_lineplot_c}) correspond to regimes $C$ and $D$ in Fig.~\ref{fig:filling_condition}(a). For both the curves, the mean error is increasing, suggesting that the threshold-$k$ filling algorithm is not suitable in these parameter regimes. It is also unlikely that these parameter regimes would be biologically relevant. 
 

 \subsection*{$k$-threshold filling model can obtain inheritance patterns similar to what is observed experimentally }

\begin{figure}

\begin{subfigure}[b]{0.5\linewidth} 
\includegraphics[width=3.5in]{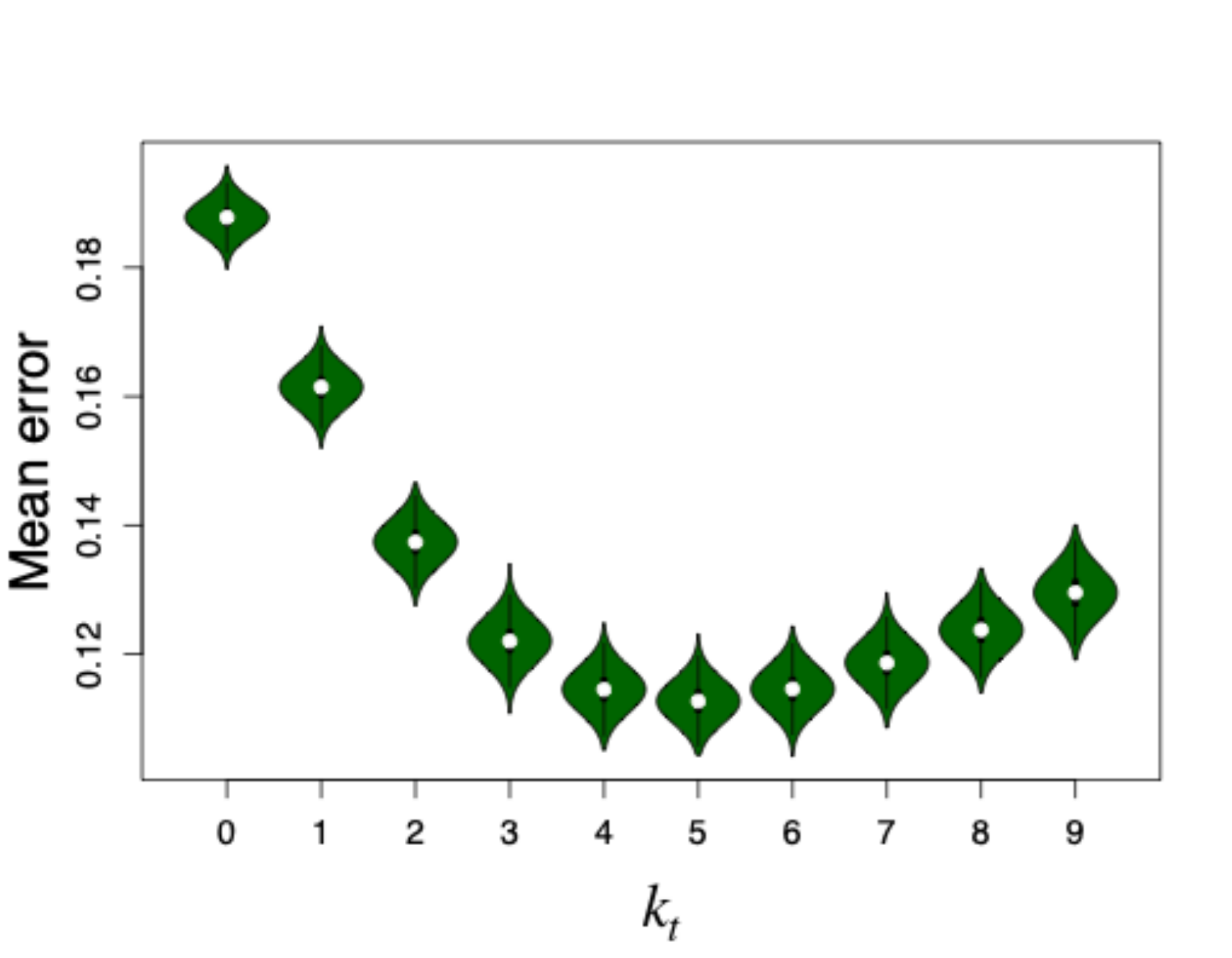}
\caption{} \label{hg_violinplot_error}
  \end{subfigure}  \vspace{0.4cm}

\begin{subfigure}[b]{0.5\linewidth}
\includegraphics[width=3.5in]{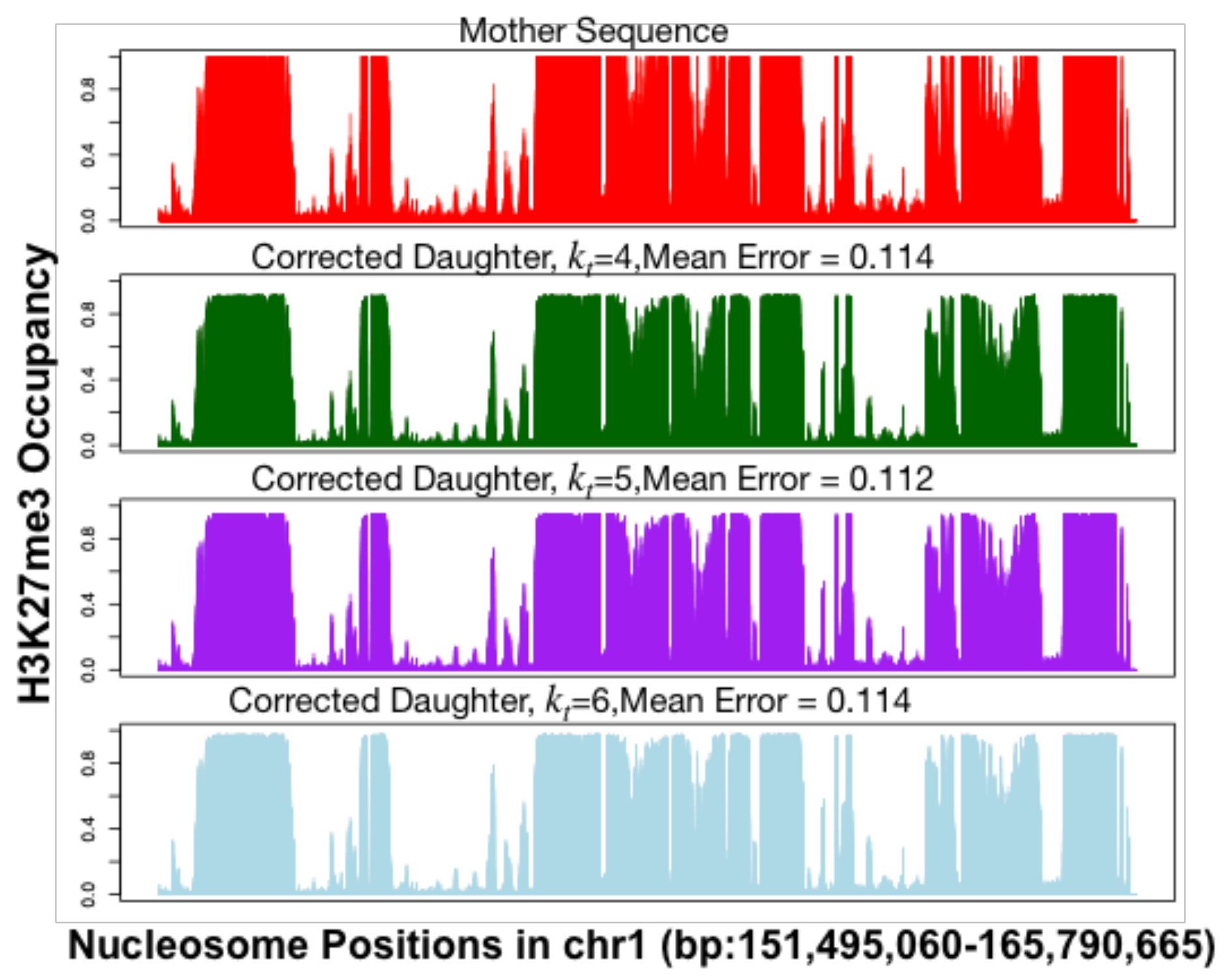}
\caption{} \label{Hg_optk_plots4}
  \end{subfigure}  

\caption{(a) Mean deviation ($\bar{\Delta}$) between corrected daughters and corresponding mothers, where the experimental population-averaged parental data for H3K27me3 is from~\cite{reveron2018accurate} (see database GEO: GSE110354).  Error correction was performed using the threshold-$k$ filling algorithm for different $k_t$ values. (b) The population averaged histone modification occupancy
 for H3K27me3 is plotted for mother sequence (top), and corrected daughter sequences corresponding to different values of $k_t$.}
\label{hg_plots}
\end{figure}

To examine the biological relevance of the findings leading to a $k$-threshold filling model, we took publicly available experimental histone modification data, and compared with our simulation results. The inheritance of modification H3K27me3  has been systematically studied recently~\cite{reveron2018accurate} by measuring modification occupancy before and after DNA replication. Since the experimental data is population-averaged, we used a simple randomized discretization algorithm to generate many binary sequences of the available data. For example, we took a long region from the chromosome 1 (151,495,060 bp to  165,790,665 bp) data, discretized to obtain the mother vector ($\bM$), and then generated the daughter chromatin $\bD = \bM \cdot \bf{Z}$. The $\bD$ was corrected to a mother-like modification sequence $\hM$ using the threshold-k algorithm for different $k=k_t$. This was repeated several times (100 $\bM$ sequences, and $100$ $\bD$ for each $\bM$), and the mean error ($\bar{\Delta}$) is plotted as a function of $k_t$ in Fig.~\ref{hg_plots}(\subref{hg_violinplot_error}).
The results show that when $k_t=5$ the mean error in the corrected H3K27me3 pattern is minimum. Note that this is very similar to the curves in Fig.~\ref{smap_optk}(\subref{opt_k_lineplot_b}) for large values of $\alpha$ and $\beta$. We independently verified that the parameters corresponding to  the original mother sequence is $\alpha \approx 0.82$ and  $\beta \approx 0.87$  implying that a biologically relevant modification falls  in parameter regime B. 

In Fig.~\ref{hg_plots}(\subref{Hg_optk_plots4}), we plot the population-averaged modification pattern for different values $k_t$. Note that islands of high and low modification occupancy regions are present in both the mother as well as the corrected daughter sequences.  This indicates that our threshold $k$-filling algorithm can reproduce biologically relevant data. Thus, our information theory-inspired algorithm predicts that there might be enzymes that simply fill short segments ($4$ or $5$ nucleosomes) of unmodified regions, but leave the longer unmodified regions ($>5$ nucleosomes) unfilled. This helps in maintaining the fidelity during epigenetic inheritance. 

It is an important question that how do enzymes know what is the optimal threshold for filling.  There are different families of enzymes, and each may have evolved differently. Some enzymes may be naturally adept at filling short regions, i.e., optimal $k$ value ($k^*$) is hard-wired into such enzymes, via evolution.  Another possibility is that other phenomena like local looping, phase separation etc. decide the threshold $k^*$, by bringing unfilled nucleosomes together~\cite{mir2019chromatin,rowley2018organizational} . These need to be understood in future.  

\subsection*{Spatially distinct antagonistic modifications}

\begin{figure}
\centering
\includegraphics[width=.6\linewidth]{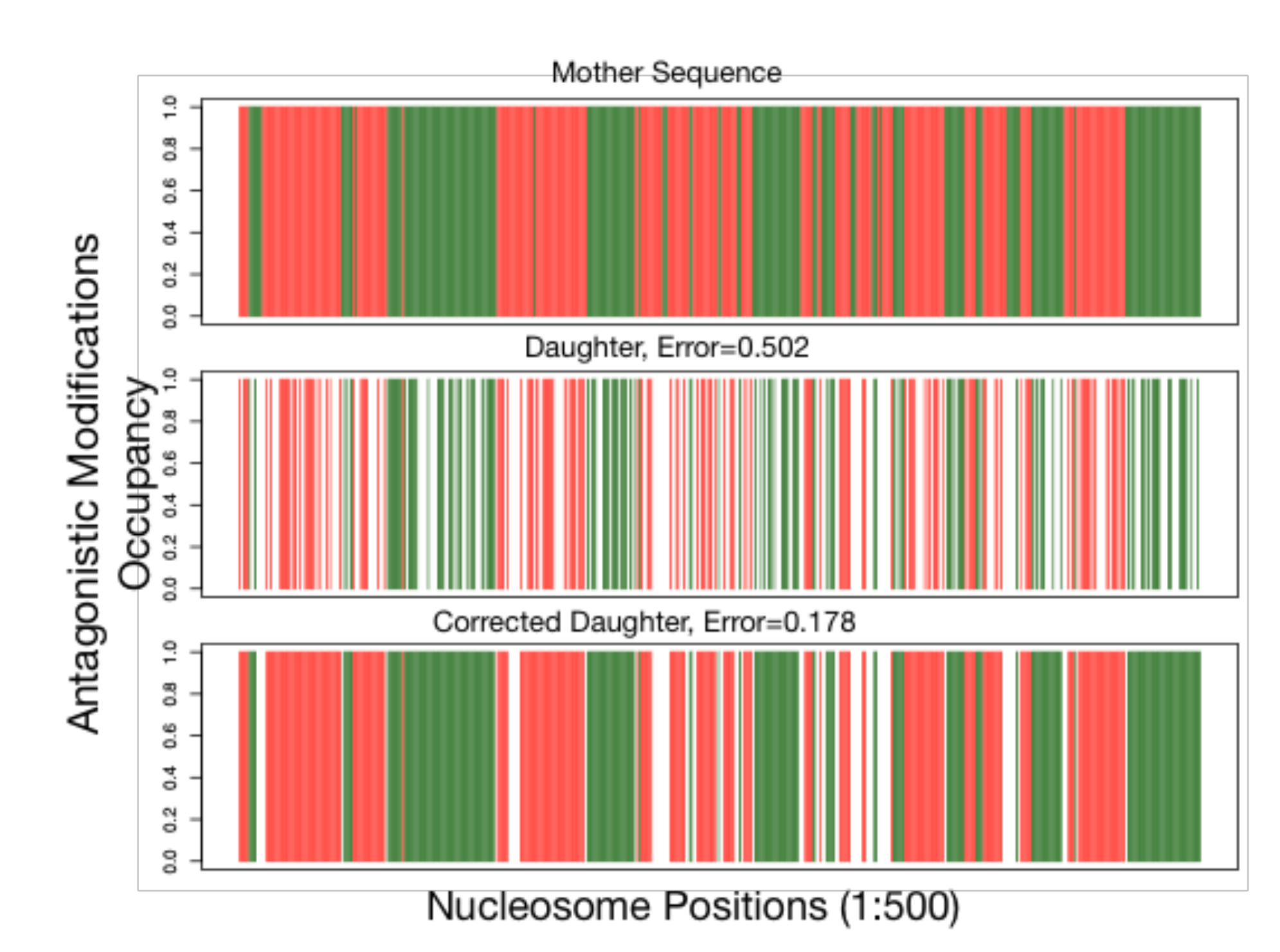}
\caption{Typical realisations of the modification patterns from the study of two spatially distinct/antagonistic modifications.  The red and green regions represent the modifications $1$ and $2$, respectively, spread over 500 nucleosomes.  Corrections were performed using the threshold-$k$ algorithm with an optimum $k_t=6$. }
\label{fig:ant_modifications_simulation}
\end{figure}

The threshold-$k$ filling model can be naturally extended to study two (or multiple) modifications that are antagonistic, spatially distinct (the same nucleosome will not have both the modifications simultaneously), and to be acted upon by very different enzymes. 
As per our model, for an enzyme-$1$ responsible for modification-$1$, the nucleosomes having the second modification are not ``visible'' and would appear as a long stretch of  $0$s. For enzyme-$2$ , similarly the modification-$1$ nucleosomes appear as a long stretch of $0$s. We generated a sequence having two spatially distinct modifications. A simple extended version of the threshold-$k$ algorithm was applied to each enzyme separately. That is, whenever there is an island of size $k\leq k_t$ between two nucleosomes having modification-$i$, the region was filled with modification-$i$, for $i=1,2$.
Anything else was left unfilled. This gave us results as shown in Fig.~\ref{fig:ant_modifications_simulation}. These are occupancies from a typical realisation (not averaged over the population) and hence the values are either $0$ or $1$.

\section*{Discussion and Conclusion}

In this work, we proposed that the problem of daughter chromatin retrieving histone modification patterns, to achieve a mother-like chromatin state, can be mapped to a communication theory problem of receiving noisy signal and correcting it to retrieve the original signal. Using ideas from Information theory, we argued that if enzymes were ideal computing machines, the best they could do is to execute a MAP decoding algorithm to get back a mother-like sequence. We  showed how well this algorithm would reconstruct the mother--- the error can be as low as 5\% in certain parameter regimes. However, the question was whether realistic enzymes can practically do such complex algorithms. We showed that in a biologically relevant parameter regime, MAP decoding algorithm is equivalent to a $k$-threshold filling algorithm. That is, the enzymes could simply insert modifications in $k$-sized or shorter unmodified stretches($0$s). The fact that a detailed theory simplifies to a process that is potentially  executable by enzymes makes this result attractive.  

We modelled the mother chromatin as a first order Markov process. This is a reasonable model as there are minimal number of parameters. One could potentially consider interactions among nucleosomes beyond the immediate neighbours. However, this will involve many more parameters, making the problem more complex and difficult to gain insights. This may be probed in a future work.

We would like to stress that given any experimental data, we do not have to know the parameters $\alpha$ and $\beta$. By simply varying $k_t$, as shown in Figs.~\ref{smap_optk}(\subref{opt_k_lineplot_b})  and \ref{hg_plots}(\subref{hg_violinplot_error}) , we can determine the optimal configurations and obtain insights about how enzymes might work. Note that even in the large $\alpha$ regime (region A in Fig.~\ref{smap_optk}(\subref{smap_plot_region})), if enzymes settle for $k$-filling with a finite $k_t$ (e.g., $5$ or $6$), it becomes a pragmatic modification correction solution as the resulting error is relatively low (see read and green curves in Fig.~\ref{smap_optk}(c)).

It is also interesting  to examine if there are noise in the insertion (filling) process itself, and how much can it affect the result. Finally, it would be also of great interest to study how the polymer nature of the chromatin would work in tune with the results from information theory. After all, the epigenetic code might involve an interplay between the one-dimensional histone codes and polymer dynamics of the chromatin. The fact that we have two $1s$ at the boundaries suggests some potential role of looping or micro phase separation in far-away regions coming together. Our own earlier studies~\cite{bajpai2020irregular} hint to us that small patches of unmodified nucleosomes (newly inserted nucleosomes) may lead to small clusters, influencing the kinetics of the modification process itself. These are questions that await future studies.

\bibliography{bsbe}{}
\bibliographystyle{plos2009}
\nocite{}

 \end{document}